\documentclass[cameraready]{Interspeech}

\let\;\thickspace
\let\|\Vert

\usepackage{amsmath}
\usepackage{amssymb}
\usepackage{booktabs}
\usepackage{multirow}
\usepackage{graphicx}
\usepackage{arydshln}

\title{Dual-Granularity Orthogonal Disentanglement for Generalizable Audio Deepfake Detection}

\author[affiliation={1}, orcid=0009-0009-9535-2941]{Zhuodong}{Liu}
\author[affiliation={1}]{Hugen}{Lv}
\author[affiliation={2}, orcid=0000-0002-4226-1255, correspondingauthor]{Xiangyu}{Li}
\author[affiliation={3}, orcid=0009-0004-6507-0496]{Chunhong}{Yuan}

\address{
    $^1$ Beijing Jiaotong University, China \\
    $^2$ Shanghai Jiao Tong University, China \\
    $^3$ ITMO University, Russia
}
\email{22711104@bjtu.edu.cn, 23722056@bjtu.edu.cn, xiangyuli@sjtu.edu.cn, 521031@niuitmo.ru}

\keywords{audio deepfake detection, disentangled representation, cross-domain generalization, orthogonality constraint}

\begin{document}

\maketitle

\begin{abstract}
Audio deepfake detectors often fail to generalize across speakers, as they learn speaker-identity features rather than synthesis artifacts, known as implicit identity leakage. 
Existing methods address this but incur architectural complexity or training instability. 
This paper proposes a dual-granularity orthogonal disentanglement framework enforcing feature independence at two levels: sample-level cosine orthogonality captures directional decorrelation, while batch-level cross-covariance regularization eliminates linear correlations across embedding dimensions.
A curriculum disentanglement schedule progressively strengthens the orthogonality constraint without auxiliary networks or adversarial dynamics.
Experiments on ASVspoof 2019 LA, ASVspoof 2021 DF, and In-the-Wild datasets demonstrate that the proposed method achieves 1.35\%, 7.88\%, and 21.58\% equal error rates (EER), respectively, surpassing gradient reversal disentanglement by 2.60\% absolute on cross-dataset transfer.
\end{abstract}
%==================================================================================
%==================================================================================
\section{Introduction}

Advances in voice conversion~\cite{sisman2020vc, huang2020vc} and text-to-speech~\cite{casanova2022yourtts, ren2020fastspeech} have enabled highly realistic synthetic speech, threatening speaker verification systems and enabling fraud and misinformation. 
While state-of-the-art detectors achieve below 2\% equal error rates (EER) on ASVspoof 2019 LA~\cite{jung2022aasist, tak2021rawnet2, wang2020asvspoof}, performance degrades to over 20\% EER on real-world data~\cite{muller2022inthewild, oneata2024generalisable}, a gap largely attributed to implicit identity leakage---detectors capturing speaker-specific characteristics rather than synthesis artifacts~\cite{ahmadiadli2025beyond, yi2023survey, wang2023asdg}.
Models exploit spurious speaker-label correlations instead of learning transferable artifact features~\cite{ xie2024domain}, as further evidenced by the ASVspoof 5 challenge~\cite{wang2025asvspoof5}, where crowdsourced recordings with severe statistical mismatch caused widespread performance degradation.

Early audio deepfake detection relied on handcrafted features such as LFCCs with GMM backends~\cite{sahidullah2015lfcc}, while recent end-to-end architectures like AASIST~\cite{jung2022aasist} and RawNet2~\cite{tak2021rawnet2} achieve strong in-domain performance.
Self-supervised representations from Wav2Vec2~\cite{baevski2020wav2vec} and WavLM further improve robustness~\cite{martin2022vicomtech, tak2022wav2vec}, with recent work demonstrating that attentive merging of hidden layers~\cite{pan2024attentive} and lightweight nested backends~\cite{liu2025nes2net} can further narrow the generalization gap.
Speaker embedding techniques such as x-vectors~\cite{snyder2018xvector} and ECAPA-TDNN~\cite{desplanques2020ecapa} have also been adapted for spoofing detection.
Despite these advances, cross-dataset evaluation reveals persistent generalization failures~\cite{muller2022inthewild, yi2023survey, combei2025unmasking} attributed to implicit identity leakage, where models memorize speaker-specific patterns rather than transferable artifacts~\cite{ahmadiadli2025beyond}.

Several disentanglement approaches have emerged to address identity leakage~\cite{ganin2016domain, bengio2013representation}.
Adversarial methods employ gradient reversal for domain-invariant representations: ASDG~\cite{wang2023asdg} targets codec invariance, while ALDEN~\cite{xu2025alden} combines adversarial training with reconstruction-based learning and meta-learning to achieve robust cross-vocoder generalization.
The Beyond Identity framework~\cite{ahmadiadli2025beyond} achieves implicit disentanglement through frequency-domain swaps and time-domain manipulations, and reconstruction-based approaches like SafeEar~\cite{li2024safeear} separate acoustic and semantic components through auxiliary generation tasks.
In the broader speech domain, mutual information minimization has been widely adopted for content-speaker disentanglement~\cite{krishna2024learn2diss, qian2022contentvec}, and orthogonal transformations have been explored for speaker anonymization~\cite{miao2023householder}.
Though being effective, these methods typically require substantial complexity: multiple encoders, auxiliary networks, MI estimators, or carefully engineered augmentation pipelines.
Orthogonality constraints have proven effective for feature independence in adjacent domains, including speech emotion recognition~\cite{zou2022dsnet} and face anti-spoofing~\cite{liu2020dsdg}, but have not been explored for speaker-artifact disentanglement in audio deepfake detection.

To address these limitations without incurring additional architectural overhead, we propose a dual-granularity orthogonal disentanglement framework.
Unlike single-level constraints that address only directional alignment, the proposed approach enforces feature independence at two complementary granularities: sample-level cosine orthogonality eliminates directional correlation between individual embedding pairs, while batch-level cross-covariance regularization, inspired by decorrelation principles in self-supervised learning~\cite{zbontar2021barlow}, removes linear correlations across embedding dimensions within each mini-batch.
A curriculum disentanglement schedule progressively increases constraint strength, informed by our empirical finding that premature aggressive disentanglement degrades performance.
This framework provides three advantages over prior disentanglement methods: deterministic optimization without adversarial instability, multi-level independence enforcement without auxiliary MI estimators, and minimal overhead allowing integration into existing pipelines.

The main contributions of this paper are as follows.
First, we introduce dual-granularity orthogonal disentanglement combining sample-level cosine orthogonality with batch-level cross-covariance regularization, representing the first multi-level geometric enforcement of speaker-artifact independence in audio deepfake detection.
Second, we propose a curriculum disentanglement schedule that progressively strengthens orthogonality constraints, preventing premature feature collapse while achieving stronger final disentanglement.
Third, we demonstrate through comprehensive experiments that this approach achieves 1.35\% and 7.88\% EER on ASVspoof 2019 LA and 2021 DF~\cite{yamagishi2021asvspoof}, respectively, and 21.58\% EER on In-the-Wild, performance comparable to self-supervised models with over 300M parameters using only 2.1M parameters, while outperforming gradient reversal adversarial training by 2.60\% absolute on cross-dataset transfer.
%==================================================================================
%==================================================================================
\section{Proposed Method}

\subsection{Problem Formulation}

Let $\mathbf{X} \in \mathbb{R}^{F \times T}$ denote the log-mel spectrogram of an input utterance with $F$ mel frequency bins and $T$ time frames.
Given a labeled training set $\mathcal{D} = \left\{(\mathbf{X}_i, y_i, s_i)\right\}_{i=1}^{N}$, where $y_i \in \{0,1\}$ indicates spoofed or bonafide and $s_i \in \{1, \ldots, K\}$ denotes speaker identity, the goal of this paper is to learn a detector that generalizes to unseen speakers by preventing implicit identity leakage, i.e., ensuring that detection relies on synthesis artifacts rather than speaker-dependent features.

We decompose the representation into a content embedding $\mathbf{z}_c \in \mathbb{R}^d$ capturing synthesis artifacts and an identity embedding $\mathbf{z}_s \in \mathbb{R}^d$ capturing speaker characteristics, and enforce orthogonality $\mathbf{z}_c \perp \mathbf{z}_s$ so that the content branch is geometrically prevented from encoding speaker information.
Let $\theta = \{\theta_{\mathrm{sh}}, \theta_c, \theta_s, \theta_{\mathrm{cls}}\}$ denote the parameters of the shared encoder, content branch, identity branch, and classification heads. The learning objective is written as:
\begin{equation}
    \min_{\theta} \; \underbrace{\mathcal{L}_{\mathrm{nat}}(\mathbf{z}_c, y)}_{\text{detection}} + \alpha \underbrace{\mathcal{L}_{\mathrm{id}}(\mathbf{z}_s, s)}_{\text{speaker supervision}} + \beta(t) \underbrace{\mathcal{L}_{\mathrm{dis}}(\mathbf{z}_c, \mathbf{z}_s)}_{\text{orthogonality}},
\label{eq:objective}
\end{equation}
where $\alpha$ controls the identity supervision strength, and $\beta(t) \in [0, \beta_{\max}]$ follows a curriculum schedule that progressively strengthens the orthogonality constraint during training (Section~\ref{sec:training}).
$\mathcal{L}_{\mathrm{id}}$ is computed on bonafide samples only ($y_i{=}1$) to learn genuine speaker characteristics without contamination from synthesis artifacts.
The specific forms of all loss terms are detailed in Sections~\ref{sec:disentangle} and~\ref{sec:training}.

\subsection{Architecture Overview}

\begin{figure}[t]
    \centering
\includegraphics[width=1.0\columnwidth]{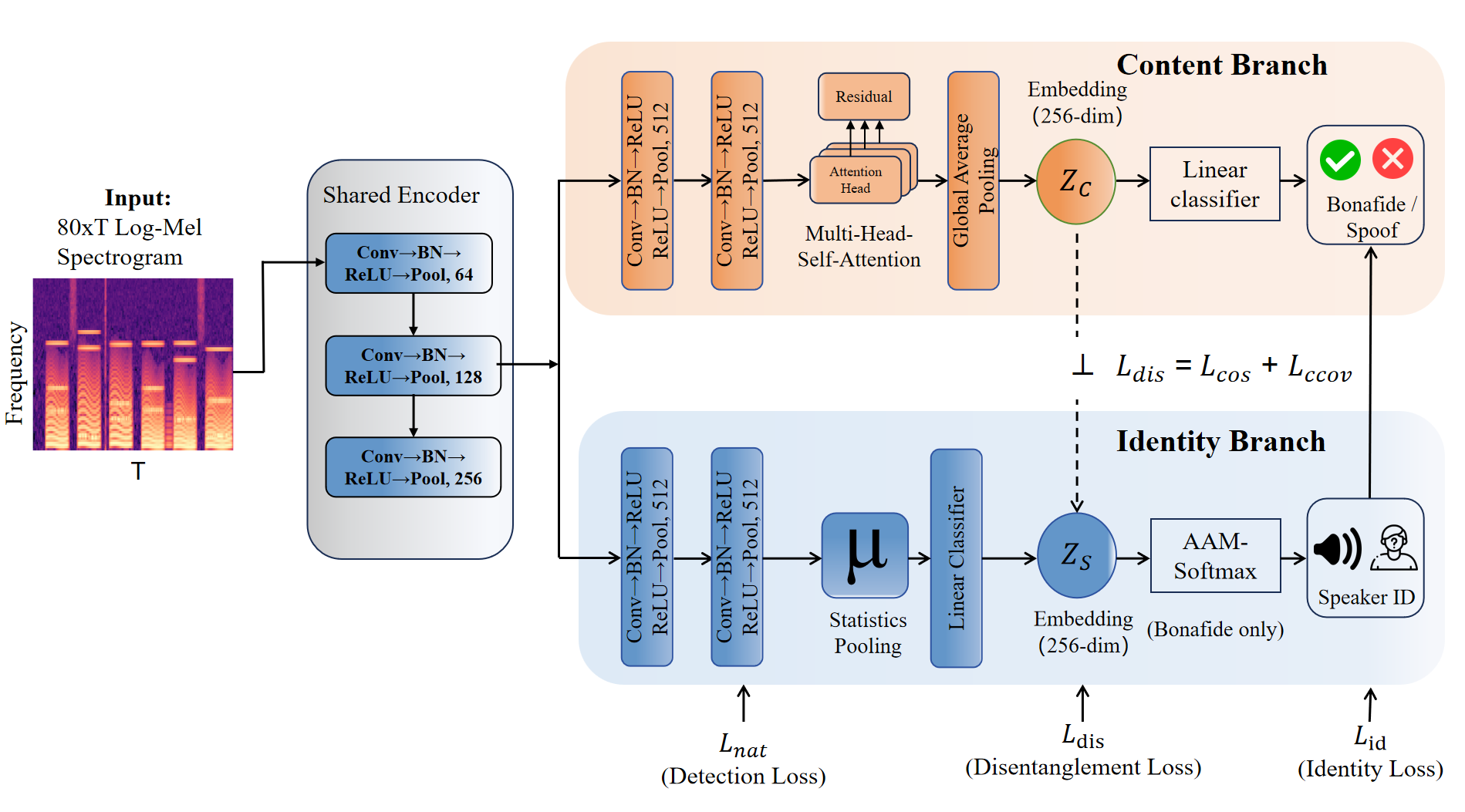} 
    \caption{Overview of the proposed dual-branch architecture with dual-granularity orthogonal disentanglement.}
    \label{fig:framework}
    \vspace{-4mm}
\end{figure}

As shown in Figure~\ref{fig:framework}, the proposed architecture comprises a shallow shared encoder $E_{\mathrm{sh}}$, a content branch $E_c$ with multi-head self-attention, and an identity branch $E_s$ with statistics pooling. The complete model contains 2.1M parameters and requires 0.89 GFLOPs per inference.

The shared encoder consists of 3 convolutional blocks ($3 \times 3$ conv, batch normalization, ReLU, $2 \times 2$ max pooling) with channels $1 \rightarrow 64 \rightarrow 128 \rightarrow 256$, reducing spatial dimensions by $8\times$. We keep this encoder shallow to preserve fine-grained acoustic details before branching.

The content branch comprises 2 additional convolutional blocks ($256 \rightarrow 512 \rightarrow 512$) followed by multi-head self-attention (MHSA) with 8 heads:
\begin{equation}
    \mathbf{H}_{\mathrm{attn}} = \mathrm{MHSA}(\mathbf{H}', \mathbf{H}', \mathbf{H}') + \mathbf{H}',
\end{equation}
where $\mathbf{H}'$ is the reshaped feature sequence from the last convolutional block, serving as queries, keys, and values in self-attention. A linear projection with global average pooling yields $\mathbf{z}_c \in \mathbb{R}^{d}$ with $d = 256$.

The identity branch shares the same convolutional configuration but replaces attention with mean statistics pooling over the frequency and time dimensions of the feature map:
\begin{equation}
    \boldsymbol{\mu} = \frac{1}{F' T'} \sum_{f,t} \mathbf{H}_{:,f,t},
\end{equation}
where $F'$ and $T'$ denote the frequency and time dimensions after downsampling. A linear projection yields $\mathbf{z}_s \in \mathbb{R}^{d}$, capturing global speaker characteristics such as vocal tract properties and prosodic patterns.

\subsection{Dual-Granularity Disentanglement}
\label{sec:disentangle}

We note that single-level disentanglement constraints are insufficient: sample-level cosine similarity captures directional alignment but ignores cross-dimensional dependencies, while batch-level statistics alone may miss per-sample entanglement.

\textbf{Sample-level cosine orthogonality.}
We minimize the absolute cosine similarity between content and identity embeddings for each sample:
\begin{equation}
    \mathcal{L}_{\mathrm{cos}} = \mathbb{E}\left[\left|\frac{\mathbf{z}_c \cdot \mathbf{z}_s}{\|\mathbf{z}_c\|\,\|\mathbf{z}_s\|}\right|\right].
\end{equation}

\textbf{Batch-level cross-covariance regularization.}
Within a mini-batch of $B$ samples, we compute mean-centered embedding matrices $\bar{\mathbf{Z}}_c, \bar{\mathbf{Z}}_s \in \mathbb{R}^{B \times d}$ and minimize the squared Frobenius norm of the cross-covariance matrix $\mathbf{C} = \frac{1}{B-1}\bar{\mathbf{Z}}_c^\top \bar{\mathbf{Z}}_s$:
\begin{equation}
    \mathcal{L}_{\mathrm{ccov}} = \frac{1}{d^2}\|\mathbf{C}\|_F^2 = \frac{1}{d^2}\sum_{i=1}^{d}\sum_{j=1}^{d}C_{ij}^2.
\end{equation}
This penalizes correlations between \emph{individual dimensions} across the two spaces, capturing fine-grained second-order dependencies (linear correlations) that cosine orthogonality alone cannot detect. Inspired by decorrelation in self-supervised learning~\cite{zbontar2021barlow}, the formulation in this paper differs by enforcing cross-space decorrelation between semantically distinct branches. The combined disentanglement loss is represented as:
\begin{equation}
    \mathcal{L}_{\mathrm{dis}} = \mathcal{L}_{\mathrm{cos}} + \gamma \, \mathcal{L}_{\mathrm{ccov}},
\end{equation}
where $\gamma = 1.0$ balances the two granularities, with robust performance observed across $\gamma \in [0.5, 2.0]$, confirming that the two constraints are complementary rather than redundant. Substituting into Eq.~(\ref{eq:objective}), the total training objective expands as:
\begin{equation}
    \mathcal{L} = \mathcal{L}_{\mathrm{nat}} + \alpha \, \mathcal{L}_{\mathrm{id}} + \beta(t) \left(\mathcal{L}_{\mathrm{cos}} + \gamma \, \mathcal{L}_{\mathrm{ccov}}\right),
\end{equation}
where $\beta(t)$ is a curriculum weight defined below.

\subsection{Training Objectives}
\label{sec:training}

The naturalness loss employs binary cross-entropy on the content embedding:
\begin{equation}
    \mathcal{L}_{\mathrm{nat}} = -\left[y \log \sigma(f_c(\mathbf{z}_c)) + (1-y) \log (1-\sigma(f_c(\mathbf{z}_c)))\right].
\end{equation}
The identity loss uses AAM-Softmax~\cite{deng2019arcface} on bonafide samples only ($y_i{=}1$):
\begin{equation}
    \mathcal{L}_{\mathrm{id}} = \mathrm{AAM}(\mathbf{z}_s, s_i;\, m, s), \quad (y_i = 1).
\end{equation}

\textbf{Curriculum disentanglement schedule.}
The analysis (Section~4.4) shows that aggressive early disentanglement degrades performance. We adopt a cosine warm-up for the disentanglement weight:
\begin{equation}
    \beta(t) = \beta_{\max} \cdot \frac{1}{2}\left(1 - \cos\left(\frac{\pi \, t}{T}\right)\right),
\end{equation}
allowing branches to first establish discriminative representations before progressively enforcing independence. While similar scheduling exists for adversarial adaptation~\cite{ganin2016domain}, our application to geometric constraints addresses a different challenge: preventing embedding collapse under orthogonality pressure rather than stabilizing minimax dynamics. The AAM-Softmax and BCE losses jointly prevent collapse by requiring discriminative representations in both branches.

%==================================================================================
\section{Experimental Setup}

We evaluate on ASVspoof 2021 DF~\cite{yamagishi2021asvspoof}, which contains bonafide speech and spoofed samples from over 100 different synthesis systems including VCC2018 and VCC2020 voice conversion submissions. The training set contains 22,617 bonafide and 22,296 spoofed utterances from 107 speakers.
For cross-dataset evaluation, we test models trained on ASVspoof 2021 DF directly on In-the-Wild~\cite{muller2022inthewild}, a challenging collection of 31,779 real-world deepfakes from diverse online sources with unknown generation methods.

All audio is resampled to 16 kHz and converted to 80-dimensional log-mel spectrograms using 512-point FFT with 160-sample hop length (10ms frame shift).
Models are trained for 50 epochs using AdamW optimizer with learning rate $10^{-4}$, weight decay $10^{-4}$, and batch size 32.
Hyperparameters are set as $\alpha = 0.1$, $\beta_{\max} = 0.5$, $\gamma = 1.0$, AAM-Softmax margin $m = 0.2$ and scale $s = 30$.
We compare against traditional methods (LFCC-GMM, LFCC-LCNN~\cite{lavrentyeva2019lcnn}), end-to-end architectures (RawNet2~\cite{tak2021rawnet2}, AASIST~\cite{jung2022aasist}), self-supervised approaches (Wav2Vec2-AASIST~\cite{tak2022wav2vec}, WavLM-MLP), and disentanglement methods including GRL adversarial training~\cite{ganin2016domain} and domain generalization via aggregation-separation~\cite{xie2024domain}.
All baselines are reimplemented under the preprocessing pipeline for fair comparison, except SSL-based methods which use their respective raw waveform frontends.

\section{Results and Discussion}

\subsection{In-Domain Detection}

Table~\ref{tab:main_results} presents detection performance on ASVspoof 2019 LA and 2021 DF. On ASVspoof 2019 LA, the proposed method achieves 1.35\% EER, ranking second among non-pretrained methods behind AASIST (0.83\%), which benefits from graph-based spectro-temporal modeling optimized for in-domain conditions. The proposed method substantially outperforms all disentanglement-based approaches, including the GRL baseline (5.30\%) and DG-Agg (1.87\%). SSL-based methods (Wav2Vec2-AASIST, WavLM-MLP) achieve lower EER, as expected, given their more than 300 M parameters and large-scale pretraining data. 
On ASVspoof 2021 DF, the proposed method achieves 7.88\% EER, comparable to WavLM-MLP (7.95\%) despite using approximately 150$\times$ fewer parameters, suggesting that explicit identity disentanglement can partially compensate for the acoustic coverage that self-supervised pretraining provides. Notably, this is achieved with 2.1M parameters and 0.89 GFLOPs, representing approximately 150$\times$ parameter reduction with less than 0.1\% absolute EER difference on ASVspoof 2021 DF, highlighting the parameter efficiency of explicit disentanglement.
DG-Agg~\cite{xie2024domain} achieves competitive in-domain performance (8.26\%) but inferior cross-dataset transfer (Table~\ref{tab:cross_dataset}), indicating that identity-level disentanglement addresses a complementary generalization bottleneck beyond domain-level aggregation.

\begin{table}[t]
    \caption{In-domain detection performance on ASVspoof 2019 LA and 2021 DF evaluation sets. Bold indicates best among methods of comparable scale ($\leq$5M parameters).}
    \label{tab:main_results}
    \centering
    \small
    \setlength{\tabcolsep}{4pt}
    \begin{tabular}{lcccc}
        \toprule
        \multirow{2}{*}{Method} & \multicolumn{2}{c}{ASV19-LA} & \multicolumn{2}{c}{ASV21-DF} \\
        \cmidrule(lr){2-3} \cmidrule(lr){4-5}
         & EER & t-DCF & EER & t-DCF \\
        \midrule
        LFCC-GMM & 8.09 & 0.2116 & 22.38 & 0.5765 \\
        LFCC-LCNN & 5.06 & 0.1000 & 15.62 & 0.4567 \\
        RawNet2 & 5.13 & 0.1175 & 15.14 & 0.4198 \\
        Res-TSSDNet & 1.64 & 0.0480 & 9.81 & 0.3105 \\
        RawGAT-ST & 1.06 & 0.0335 & 10.75 & 0.3218 \\
        AASIST & 0.83 & 0.0275 & 12.83 & 0.3624 \\
        \hdashline
        Wav2Vec2-AASIST$^\dagger$ & 0.52 & 0.0165 & 8.54 & 0.2876 \\
        WavLM-MLP$^\dagger$ & 0.43 & 0.0148 & 7.95 & 0.2708 \\
        \midrule
        LCNN Baseline & 5.00 & 0.0510 & 15.30 & 0.4482 \\
        GRL Baseline & 5.30 & 0.0670 & 8.91 & 0.2953 \\
        DG-Agg~\cite{xie2024domain} & 1.87 & 0.0382 & 8.26 & 0.2814 \\
        Cosine Only & 1.50 & 0.0220 & 8.23 & 0.2762 \\
        \textbf{Full Model (Proposed)} & \textbf{1.35} & \textbf{0.0208} & \textbf{7.88} & \textbf{0.2689} \\
        \bottomrule
        \multicolumn{5}{l}{\footnotesize $^\dagger$ SSL-based methods with $>$300M parameters.}
    \end{tabular}
        \vspace{-6mm}
\end{table}

\subsection{Cross-Dataset Generalization}

Table~\ref{tab:cross_dataset} evaluates cross-dataset transfer from ASVspoof 2021 DF to In-the-Wild without fine-tuning~\cite{oneata2024generalisable, combei2025unmasking}. The proposed method achieves 21.58\% EER, comparable to WavLM-MLP (21.85\%) while outperforming the GRL baseline by 2.60\% absolute under identical architecture, confirming that multi-granularity geometric constraints provide more robust disentanglement than adversarial training. DG-Agg~\cite{xie2024domain} achieves 22.73\%, suggesting that identity-level and domain-level generalization address complementary aspects of distribution shift.
The progression from single-branch (25.86\%) to no-disentangle (23.48\%) to cosine-only (22.16\%) to the full model (21.58\%) demonstrates cumulative benefits at each stage of framework.

\begin{table}[t]
    \caption{Cross-dataset generalization. Models trained on ASVspoof 2021 DF and evaluated on In-the-Wild without fine-tuning.}
    \label{tab:cross_dataset}
    \centering
    \small
    \begin{tabular}{lcc}
        \toprule
        Method & ASV21-DF (\%) & ITW (\%) \\
        \midrule
        AASIST & 12.83 & 27.41 \\
        Res-TSSDNet & 9.81 & 26.14 \\
        RawGAT-ST & 10.75 & 25.83 \\
        Wav2Vec2-AASIST$^\dagger$ & 8.54 & 23.17 \\
        WavLM-MLP$^\dagger$ & 7.95 & 21.85 \\
        GRL Baseline & 8.91 & 24.18 \\
        DG-Agg~\cite{xie2024domain} & 8.26 & 22.73 \\
        \midrule
        Single Branch & 10.17 & 25.86 \\
        No Disentangle & 9.05 & 23.48 \\
        Cosine Only & 8.23 & 22.16 \\
        \textbf{Full Model (Proposed)} & \textbf{7.88} & \textbf{21.58} \\
        \bottomrule
        \multicolumn{3}{l}{\footnotesize $^\dagger$ SSL-based methods with $>$300M parameters.}
    \end{tabular}
        \vspace{-4mm}
\end{table}

We note that ALDEN~\cite{xu2025alden} and Beyond Identity~\cite{ahmadiadli2025beyond} represent closely related disentanglement approaches; however, direct numerical comparison is not straightforward as ALDEN evaluates primarily on cross-vocoder settings with different data splits, and Beyond Identity employs task-specific augmentation pipelines with distinct evaluation protocols. Both methods require substantially more architectural complexity (auxiliary reconstruction networks and meta-learning for ALDEN; multi-stage augmentation for Beyond Identity), whereas the proposed approach achieves competitive generalization through two additional loss terms alone.

\subsection{Ablation Study}

Table~\ref{tab:ablation} quantifies each component's contribution on ASVspoof 2019 LA. The identity branch and AAM-Softmax cause the largest degradations when removed ($+4.30\%$ and $+4.04\%$), confirming that explicit speaker modeling is essential for the content branch to focus on artifacts. Among disentanglement components, cosine orthogonality contributes more than cross-covariance alone (1.85\% vs.\ 2.38\% EER), yet combining both yields 1.35\%, confirming complementary coverage of directional and cross-dimensional dependencies. The curriculum schedule further improves over fixed $\beta$ by $+0.38\%$, validating the progressive strengthening strategy.

\begin{table}[t]
    \caption{Ablation study on ASVspoof 2019 LA evaluation set.}
    \label{tab:ablation}
    \centering
    \small
    \begin{tabular}{lcc}
        \toprule
        Configuration & EER (\%) & $\Delta$ EER \\
        \midrule
        Full Model & 1.35 & -- \\
        \quad w/o Identity Branch & 5.65 & +4.30 \\
        \quad w/o $\mathcal{L}_{\mathrm{cos}}$ \& $\mathcal{L}_{\mathrm{ccov}}$ & 2.84 & +1.49 \\
        \quad w/o $\mathcal{L}_{\mathrm{ccov}}$ (cosine only) & 1.85 & +0.50 \\
        \quad w/o $\mathcal{L}_{\mathrm{cos}}$ (ccov only) & 2.38 & +1.03 \\
        \quad w/o Curriculum (fixed $\beta$) & 1.73 & +0.38 \\
        \quad w/o AAM-Softmax & 5.39 & +4.04 \\
        \quad w/o Self-Attention & 3.52 & +2.17 \\
        \bottomrule
    \end{tabular}
        \vspace{-4mm}
\end{table}

\subsection{Sensitivity and Disentanglement Analysis}

\begin{figure}[t]
    \centering
    \includegraphics[width=0.80\columnwidth]{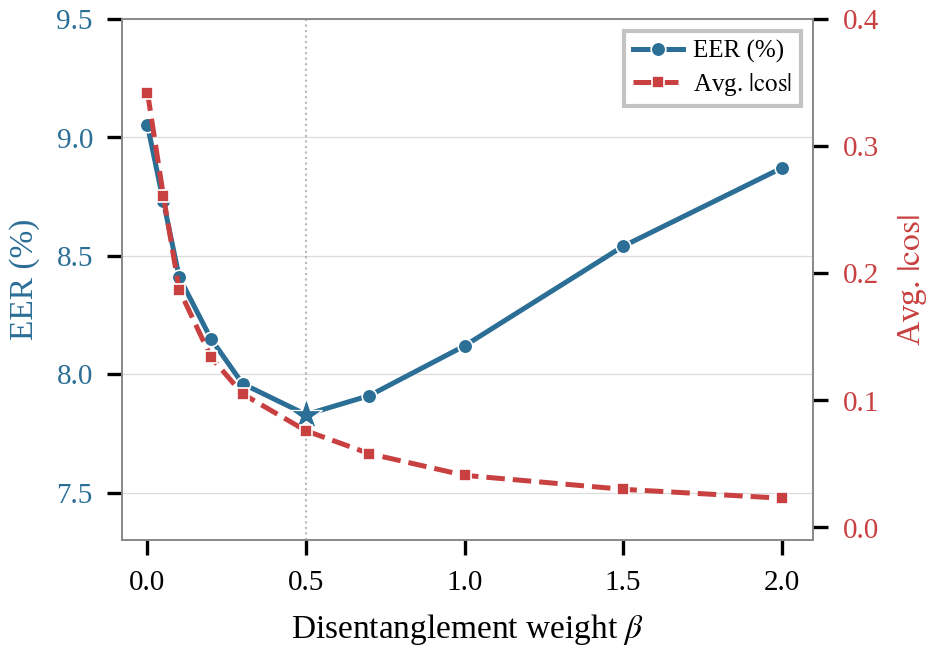}
    \caption{Sensitivity of EER and cosine similarity to $\beta_{\max}$ on ASVspoof 2021 DF. Dual-granularity with curriculum (solid) vs.\ cosine-only with fixed $\beta$ (dashed).}
    \label{fig:beta}
    \vspace{-4mm}
\end{figure}

Figure~\ref{fig:beta} shows the sensitivity to $\beta_{\max}$ on ASVspoof 2021 DF. Without orthogonality ($\beta_{\max}=0$), the average cosine similarity between embeddings reaches 0.342, indicating substantial information overlap between the two branches. As $\beta_{\max}$ increases, cosine similarity decreases monotonically, confirming effective orthogonality enforcement. Notably, dual-granularity disentanglement with curriculum scheduling broadens the optimal range to $\beta_{\max} \in [0.3, 0.8]$, compared to a narrow peak at $\beta=0.5$ under cosine-only with fixed weighting, substantially improving robustness to hyperparameter selection. The optimal $\beta_{\max} = 0.5$ achieves 0.048 cosine similarity and 7.88\% EER. Per-attack analysis on ASVspoof 2019 LA further shows below 2\% EER on 12 of 13 conditions, with A17 (waveform concatenation) the most challenging at 3.41\%.

\begin{figure}[t]
    \centering
    \begin{minipage}[b]{0.48\columnwidth}
        \centering
        \includegraphics[width=\linewidth]{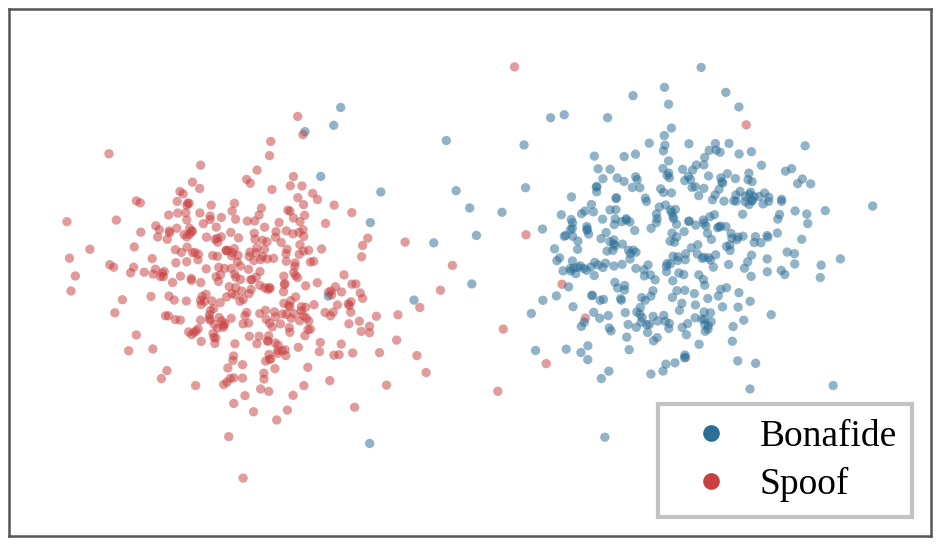}
        \centerline{(a) Colored by authenticity}
    \end{minipage}
    \hfill
    \begin{minipage}[b]{0.48\columnwidth}
        \centering
        \includegraphics[width=\linewidth]{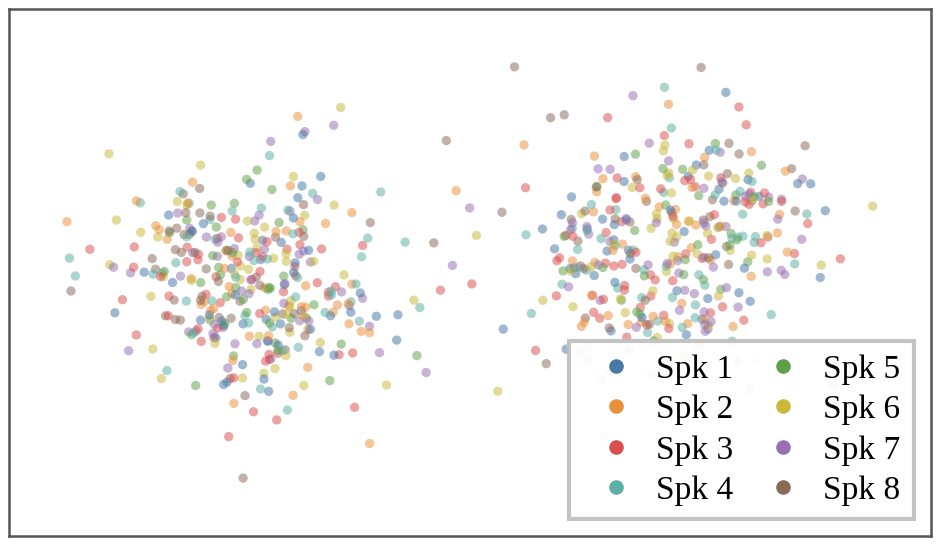}
        \centerline{(b) Colored by speaker identity}
    \end{minipage}
    \caption{t-SNE visualization of content embeddings $\mathbf{z}_c$ on In-the-Wild.}
    \label{fig:tsne}
    \vspace{-8mm}
\end{figure}

Figure~\ref{fig:tsne} visualizes content embeddings on In-the-Wild via t-SNE to qualitatively validate the disentanglement effect. Embeddings form clearly separable clusters when colored by authenticity (a), confirming discriminative power on unseen data. Crucially, the same embeddings show no discernible speaker-level structure when colored by speaker identity (b), with all speakers uniformly distributed across both clusters. This confirms that the orthogonality constraint effectively eliminates identity leakage from the content branch.
Leave-one-speaker-out cross-validation (10 folds over 107 speakers) provides further quantitative evidence: the proposed method yields a mean EER of 9.87\% $\pm$ 1.24\%, compared to cosine-only at 10.52\% $\pm$ 1.53\% and AASIST at 14.82\% $\pm$ 3.21\%. The lower mean and substantially reduced variance confirm that multi-granularity disentanglement produces more consistent performance across speaker configurations, reducing speaker-dependent overfitting beyond what single-level constraints achieve.

%==================================================================================

\section{Conclusion}

This paper presented a dual-granularity orthogonal disentanglement framework combining sample-level cosine orthogonality with batch-level cross-covariance regularization under a curriculum schedule, enforcing speaker-artifact separation without auxiliary networks or adversarial training.
This lightweight approach (2.1M parameters) achieves competitive performance with self-supervised models while outperforming adversarial disentanglement by 2.60\% absolute on cross-dataset transfer.
Future work includes extending to codec variability, controlled-variable validation of identity leakage, and integration with self-supervised frontends.
%==================================================================================
\section{Generative AI Use Disclosure}
Generative AI tools were used for English language polishing and \LaTeX{} formatting assistance during manuscript preparation. All scientific content, experimental design, implementation, analysis, and conclusions are the sole work of the authors.
%==================================================================================

\bibliographystyle{IEEEtran}
\bibliography{references}

\end{document}